\documentclass[a4paper,11pt]{article}
\usepackage{jinstpub}
\usepackage{graphicx}
\usepackage{amsmath}
\usepackage{hyperref }
\usepackage{lineno}
\usepackage{xcolor}
\usepackage{gensymb}
\usepackage{makecell}
\usepackage{hhline}
\usepackage{multirow}
\usepackage{subcaption}
\usepackage{ragged2e}
\DeclareCaptionJustification{justified}{\justifying}
\graphicspath{ {images/} }
\title{A Hybrid 3D/2D Field Response Calculation for Liquid Argon 
Detectors with PCB Based Anode Plane}

\author[a,1]{S. Martynenko\note{Corresponding author.}} 

\author[b,c]{F. Pietropaolo}

\author[a]{B. Viren}
\author[a]{X. Qian}
\author[a]{H. Chen}
\author[a]{S. Gao}
\author[a]{W. Gu}
\author[a]{J. Jo}
\author[a]{S. Kettell}
\author[a]{Y. Li}
\author[a]{H. Liu}
\author[a]{N. Nayak}
\author[a]{B. Yu}
\author[a]{H. Yu}
\author[a]{C. Zhang}
\author[b]{U. Kose}
\author[b]{F. Resnati}
\author[b]{S. Tufanli}
\author[b,d]{F. Boran}
\author[b,d]{and F. Dolek}

\affiliation[a]{Brookhaven National Laboratory, Upton, NY 11973, USA}
\affiliation[b]{CERN, CH-1211 Geneva 23, Switzerland}
\affiliation[c]{INFN, Sezione di Padova, I-35131 Padova, Italy}
\affiliation[d]{Beykent University, Istanbul, Turkey}

\emailAdd{smartynen@bnl.gov}

\abstract {Liquid Argon Time Projection Chamber (LArTPC) technology is commonly utilized in neutrino detector designs. It enables detailed reconstruction of neutrino events with high spatial precision and low energy threshold. Its field response (FR) model describes the time-dependent electric currents induced in the anode-plane electrodes when ionization electrons drift nearby. An accurate and precise FR is a crucial input to LArTPC detector simulations and charge reconstruction. Established LArTPC designs have been based on parallel wire planes. It allows accurate and computationally economic two-dimensional (2D) FR models utilizing the translational symmetry along the direction of the wires. Recently, novel LArTPC designs utilize electrodes formed on printed circuit board (PCB) in the shape of strips with through holes. The translational symmetry is no longer a good approximation near the electrodes and a new FR calculation that employs regions with three dimensions (3D) has been developed. Extending the 2D models to 3D would be computationally expensive. Fortuitously, the nature of strips with through holes allows for a computationally economic approach based on the finite-difference method (FDM). In this paper, we present a new software package {\it pochoir} that calculates LArTPC field response for these new strip-based anode designs. This package combines 3D calculations in the volume near the electrodes with 2D far-field solutions to achieve fast and precise field response computation. We apply the resulting FR to simulate and reconstruct samples of cosmic-ray muons and $^{39}$Ar decays from a Vertical Drift (VD) detector prototype operated at CERN. We find the difference between real and simulated data within 5\%. Current state-of-the-art LArTPC software requires a 2D FR which we provide by averaging over one dimension and estimate that variations lost in this average are smaller than 7\%. }


\begin{document}
\maketitle

\section{Introduction}

In past decades, the Liquid Argon Time Projection Chamber (LArTPC) technology~\cite{original_tpc} has 
become one of the main neutrino detection methods. Its high spatial precision and low energy threshold 
allow for accurate topological and energy reconstructions of the neutrino interactions. 
LArTPCs are utilized in many accelerator neutrino experiments~\cite{DUNE,SBND,microboone,lariat}. 

Charged particles traversing the argon leave behind trails of ionization electrons.
These electrons will drift in an applied electric field toward readout planes consisting of many parallel anode electrodes.
The signals acquired from the readout of these electrodes can be modeled as a convolution of the distribution of the ionization electrons, the field response (FR) describing the time-dependent electrical current induced by each electron in each electrode, and the response to this current by the amplifiers and analog-to-digital converters comprising the detector readout electronics.   Additionally, intrinsic thermal fluctuations in the electronics and any unmitigated sources of radio-frequency emission will contribute noise components.

A detailed FR is an essential input required to accurately simulate the readout of these detectors.
Likewise, an FR that is averaged over the region around each electrode is a core ingredient to a signal-processing
stage that deconvolves the sampled waveforms to produce an estimate of the original ionization electron distribution. 
This estimate is then the main ingredient to all high-level pattern recognition software that categorizes the data in terms of its event topology, particle identification, energy estimate and particle flow~\cite{Marshall:2017svj,microbooneLARP2,microbooneLARP1,MicroBooNE:2021ojx,MicroBooNE:2020sar}.
Therefore, accurate and precise field response models are essential in allowing LArTPCs to perform precision physics measurements. 

\begin{figure}[t]
    \centering
    \begin{subfigure}[b]{0.49\textwidth}
        \centering
        \includegraphics[width=1.0\textwidth]{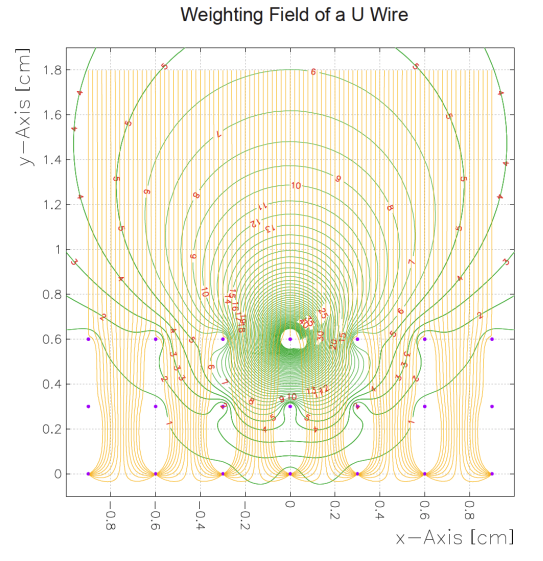}
        \caption{\centering Wire-based anode}
    \label{fig:wba}
    \end{subfigure}
\hfill    
    \begin{subfigure}[b]{0.49\textwidth}
        \centering
        \includegraphics[width=0.85\textwidth]{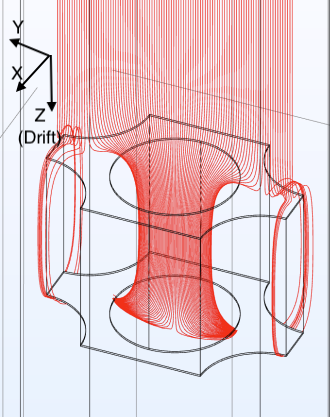}
        \caption{\centering PCB-based anode}
        \label{fig:pcbba}
    \end{subfigure}
 
    \caption{The left figure (a) shows equipotential lines of the MicroBooNE U plane Ramo weighting field (green) [Sec. \ref{sec:fralg}] and electron drifting paths (yellow) calculated for a 2D model of three wire planes (purple dots) using GARFIELD~\cite{VEENHOF1998726}.  The right figure (b) shows the electron drift paths (red) passing through holes in strips formed on PCB \cite{pcbdraft}.  Due to the applied bias potentials, paths pass through the hole in the top strip and terminate on bottom strip.  This small volume, local field calculation utilizes the finite-element method in a 3D domain. }
    \label{fig:ba}
\end{figure}

A field response model depends strongly on the design of the anode electrodes and their electrical potentials in addition to the nominal drift electric field.
Most of the recent large LArTPC detectors, including ICARUS \cite{ICARUS}, MicroBooNE \cite{microboone}, ProtoDUNE~\cite{DUNE:2020cqd}, and SBND \cite{SBND}, utilize three planes of parallel wires each providing one projective view of the ionization electrons.
A wire-based design is also chosen for the first 10 kt module of the DUNE far detector \cite{DUNE}. 
Modeling the FR of a wire-based design benefits from an approximate translational invariance along the lengths of their wires.
This allows two-dimensional (2D) FR models (Fig. \ref{fig:wba}) to be formed with high precision and accuracy while requiring a modest amount of computation.
Such models are well validated against real detector data~\cite{microbooneSPPart1}.

Recently, novel anode designs are under development that are based on parallel strip electrodes (Fig. \ref{fig:pcbba}).
Their strips are formed on printed circuit board (PCB) and pierced by through holes\cite{pcbdraft}.
One such design forms the basis for the second 10 kt module of the DUNE far detector \cite{dunecdrdraft}.
Production of PCB-based anodes is expected to be more cost-effective compared to wire-based designs and provide other advantages such as robustness in shipping and installation.
The through holes are required so the large PCB planes do not impede the natural liquid argon flow during initial filling and ongoing convection.
The holes also break the approximate translation symmetry present in wire-based LArTPCs.
This leads to significant variations in the fields as a function of
position along a strip.
The fields also approach translationaly symmetric forms some distance from the PCB plane.
Thus to accurately model the FR for these promising new designs a new method is required that spans three dimensions (3D) for at least the field region near the electrodes.

Historically, the field response calculation for the wire readout is performed using the GARFIELD package~\cite{VEENHOF1998726} that implements the finite-element method on a 2D domain.
An attempt was made to apply the COMSOL package~\cite{multiphysics1998introduction} over a 3D to a strip design.
While accurate, it was computationally intensive, particularly when modeling the long-range and position-dependent induced current~\cite{MicroBooNE:2018swd}. 
A sizable simulated volume with many incident electron positions is required and a new approach was needed.
In this paper, we present a precise, fast, and easy-to-use python package ({\it pochoir})~\cite{pochoir} that provides a hybrid 3D/2D field response calculation on CPU and GPU based on the simpler finite-difference method (FDM).
Furthermore, we validate these calculated field response models by applying the detector simulation and signal processing components of the Wire-Cell Toolkit (WCT) \cite{wcweb}.
We further validate against data taken from the 50-L Vertical Drift prototype detector operated at CERN (CERN 50-L)~\cite{dunecdrdraft}.
We utilize samples of cosmic-ray muons and $^{39}$Ar decays produced both by the simulation and acquired from the real CERN 50-L detector.

This paper is organized as follows. Sec.~\ref{sec:detector} briefly describes the  CERN 50-L detector. Sec.~\ref{sec:fralg} describes the field response calculation approach used in the
{\it pochoir} package. Sec.~\ref{sec:valid} focuses on validating the field response calculation results through a
data-simulation comparison for the CERN 50-L prototype. Finally, the results are summarized and discussed in
Sec.~\ref{sec:Summ}.

\section{Detector}\label{sec:detector}

The CERN 50-L detector was used to perform simulation tests and validations of our field response models.
This detector was a part of a large-scale test of the single-phase vertical drift technology for 
the DUNE experiment at CERN \cite{dunecdrdraft}. The data used in this paper was acquired on May 26th, 2020. 

\begin{figure}[ht]
  \centering
  \includegraphics[width=0.8\linewidth]{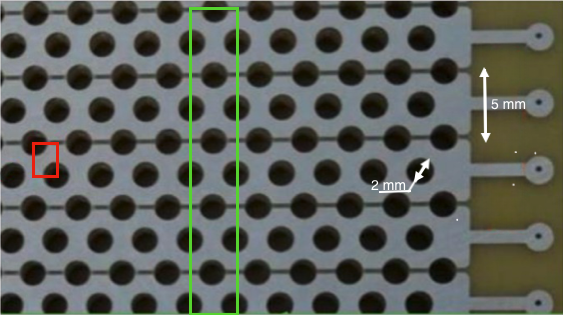}  
    \caption{Example of induction PCB plane used in CERN 50-L prototype detector with 2-mm diameter holes. 
    The red square represents minimal area used for electric drift field calculation after considering symmetries,
    while the green square shows part of minimal area for weighting field calculation after considering symmtries. 
    The total volume for weighting field calculation includes 21 strips.}
    \label{fig:strips}
\end{figure}

The detector was configured with a total drift distance of 52 cm and a uniform drift electric field of 500 V/cm. 
Waveform data was acquired from the per-strip electronics channels over a readout time of $323~\mu$s at a sampling period of 0.5$~\mu$s and in 12 bits of ADC resolution.
The pressure in the cryostat and corresponding LAr temperature was set at $\sim 100$~mbar and $\sim 87.5$~K, respectively. 
The anode electrodes were printed on a 32~cm $\times$ 32~cm $\times$ 3.2~mm double-layer PCB plate (Fig. \ref{fig:strips}). One side provided an ``induction'' layer at 0~V electric potential and facing the drift volume.
The other provided a ``collection'' layer at 2~kV and located 3.2 mm away from the induction layer, on the side of the PCB facing away from the main drift volume.
Both layers consisted of 64 readout electrode strips running in mutually orthogonal directions. 
Strips were separated by 5 mm pitch and were separated by a 0.5 mm gap.
The design utilized two different hole patterns.
The collection plane's first 32 channels (strips) had 2 mm diameter holes, while channels 32 through 64 had 2.5 mm diameter holes.

Several hardware features contributed to different charge/field responses seen by different planes.
First, different hole sizes imply different nearby electron drift paths. 
This effect is expected to be small and is discussed in Sec.~\ref{sec:valid}.
Second, capacitors were installed on the anode readout planes to separate the bias voltages from the readout electronics.
They introduced a $\sim 1\%$ contribution to the charge difference between collection and induction planes.

\section{Field Response Computation Algorithm}
\label{sec:fralg}

Utilizing the example of the field response calculation for the  CERN 50-L prototype detector, we provide 
a description of the computation approach used in the {\it pochoir} package in this section. 
The core of the field response calculation is the Ramo theorem \cite{ramo1939currents,shockley1938currents}.
It calculates the electric current $I$ induced in an electrode due to the movement of an electric charge as:
\begin{equation}
\label{eq:SRtheorem}
    I = -q\cdot \Vec{v}\cdot \Vec{E}_w,
\end{equation}
where $q$ is the electric charge, $\Vec{v}$ is its instantaneous velocity vector, and $\Vec{E}_w$ is a ``weighting'' field for the \textit{electrode of interest} (\textit{eg} one strip).
The weighting field is the electrostatic field that would result if the electrode of interest is placed at an electric potential of +1 V and all other electrodes are placed at 0 V.

The FR calculation performed by \textit{pochoir} progresses initially through three major steps: i) drift field calculation, ii) weighting field calculation, and iii) calculation of electron velocity for steps along the drift path. 
Finally, all steps are combined to calculate the induced current via Eq.~\eqref{eq:SRtheorem}.

In the first two steps, the drift and weighting fields are calculated by solving the Laplace equation for the scalar potential $\phi$ given boundary conditions specific to each field:
\begin{equation}
\label{eq:maxwell}
    \nabla^2 \phi = 0.
\end{equation}
These conditions and how the divergence of the potentials are formed are described below.

There are three well known methods to solve Eq.~\eqref{eq:maxwell}: finite difference method (FDM), finite element method (FEM), and boundary element method (BEM).
The BEM solves the boundary field on the surface mesh and then uses that boundary solution to calculate the field at any arbitrary point in space.
Thus, it requires only a surface (instead of a volume) mesh.
The FEM requires a volumetric mesh, but it can be nonuniform and ``adaptive'' so that a finer mesh can be used near detailed features and a coarser mesh can be used in the bulk volume.
Among these three methods, the FDM algorithm is the simplest to implement but requires a volumetric mesh which must be a uniform square/cubic grid.
The advantage of a uniform grid is that the method has equal precision across the whole volume.
The disadvantage is that a grid size must be chosen comparable to the smallest significant electrode feature.
FDM would lead to prohibitively costly computation for wire-based geometries but the feature size and regularity of the strip with through hole designs makes FDM feasible.
In {\it pochoir}, the FDM method was chosen
for its simplicity and the ability to execute its computations in highly parallel manner with GPU accelerators.

\begin{figure}[t]
  \centering
  \includegraphics[width=0.8\linewidth]{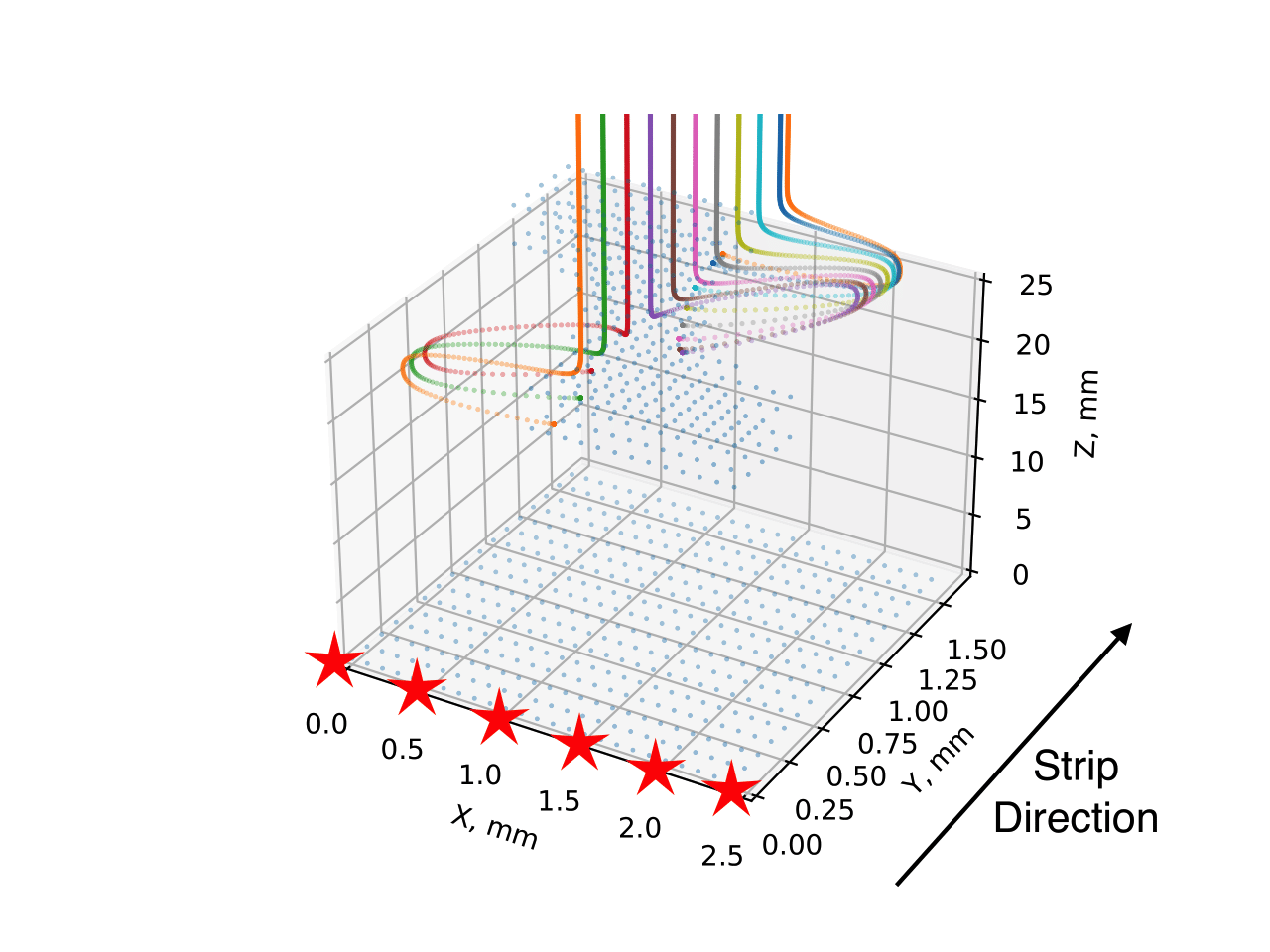}  
    \caption{The example of electron drifting paths calculated along the drifting direction as electrons passing 
    through the PCB holes. Light blue dots represent anode and volume boundaries, while colored dots are electron 
    drifting paths. Red stars represent approximate positions used to calculate the average field response for the WireCell Toolkit
    detector simulation. For the signal processing step, these field response functions are further averaged.}
    \label{fig:Path}
\end{figure}

The main parameter governing the FDM is the spacing of its uniform grid.
The need for the grid to sufficiently resolve the electrode shapes and to span a large volume calls for a fine grid spacing.
This is tempered by the need to minimize the number of grid points in order to control computational costs.
The two layers of electrodes in the CERN 50-L design (see Fig~\ref{fig:Path} and section~\ref{sec:detector}) pose another, discrete constraint.
At a minimum and neglecting electrode thickness, a plane of grid points must be chosen coplanar with each electrode layer.
This defines a maximum allowed grid spacing and requires an integral number of points between the two layers.
A yet smaller spacing is required to resolve the hole circles and to allow variation at a somewhat smaller scale than required by the applications of the final FR.
To balance between these constraints, a 0.1 mm grid size was chosen.

It is worth noting that final PCB-based designs are expected to include a second board providing two more layers: one serving as a second ``induction'' layer and an outermost passive layer providing shielding from long-range induction effects. 
This design will pose two discrete co-planar constraints that together require the ratio of board separation to board thickness to be well approximated by a simple fraction. 
The numerator and denominator of this simple fraction give an allowed number of points spanning the inter-PCB distance and PCB thickness, respectively.
The allowed error in this approximation must be small compared to the thickness of the electrodes.
The precise design choices for these three distances will determine the flexibility allowed in selecting a suitable grid spacing when applying this method.

With a discrete grid defined, the FDM assigns a value for the potential on a grid point at iteration $t+1$ as the average of values of neighboring points at iteration $t$:
\begin{equation}
\label{eq:potential}
    V^{t+1}_{i,j,k}=\frac{ \sum_{a=i\pm1,b=j\pm1,c=k\pm1}V^t_{a,b,c}}{6},
\end{equation}
where $i,j,k$ indices represent the point on the volumetric grid and indexes $a,b,c$ represent adjacent grid points along each orthogonal axis.
Note the chosen grid spacing becomes implicit.
Given definitions of boundaries and their strategy (eg fixed or periodic), boundary conditions are asserted by explicitly overwriting values following the completion of each iteration.
Convergence is reached when the maximum difference of values between iterations and across the grid points is less than a set precision.
We define this precision threshold to be $10^{-6}~$V.
The electric field at any point may then be estimated by interpolating the electric potential between nearby grid points prior to taking the gradient.


To match the dimensions of the CERN 50-L detector, all field calculations nominally assume a volume spanning 
35~mm~$\times$~3.4~mm~$\times$~210~mm.
A model for the PCB was positioned at the height of 24~mm and defined with 2.5~mm through holes.
This results in the nominal drift direction being ``downward'' toward the origin of the Z-axis in the figures.

The drift field is calculated on a smaller domain defined by exploiting periodic symmetry in the PCB design as spanning 2.5~mm~$\times$~1.7~mm~$\times$~210~mm.
The domain of this calculation in the plane of the PCB is indicated by the red rectangle in Fig.~\ref{fig:strips}.
Periodic boundary conditions are asserted along these edges.
Fixed boundary conditions representing applied potentials are asserted on the two remaining sides and on points representing electrode material.
Results inside this reduced volume are later tiled across the plane of the PCB.

The weighting field does not share the same symmetries as the drift field.
Its domain must cover multiple strips as shown in part as the green rectangle in Fig.~\ref{fig:strips}.
A total of 21 strips are included to capture long-range induction effects that exist in the direction transverse to the central strip of interest.
The choice to use 21 strips follows that same choice made in Ref.~\cite{MicroBooNE:2018swd}. 
The CERN 50-L has a larger 5-mm pitch and uses strips while MicroBooNE uses 3~mm pitch wires.
Both variations serve to reduce long range induction effects in the former relative to the latter.
Fewer strips would suffice for comparable accuracy while reducing calculation time but to keep data format consistency, the same span was chosen.
A symmetry along the strip direction remains and allows the domain of the calculation to span one period of the hole pattern by applying periodic boundary conditions to the sides perpendicular to strip direction.
In keeping with the definition of the weighting field, the central strip is fixed at 1~V and all others are fixed at 0~V.
The remaining boundaries are left free with the exception of an internal boundary that arises from the optimization described next.

Applying the FDM to calculate the weighting field over the full 3D domain as just described requires prohibitive computation resources. 
Instead, the calculation of the weighting field is comprised of two simpler calculations.
First, a 2D domain is defined as a slice of dimension 105~mm~$\times$~210~mm that cuts across the entire 3D domain and is perpendicular to the direction of the strips.
A weighting potential is calculated on this domain and then extended back along the strip direction to produce a 3D result, though degenerate in the direction of the strips.
Next, a smaller 3D domain is defined within the full domain and located near to and spanning the central seven strips.
The five sides of this smaller 3D domain are constrained to match the potentials at their same locations in the larger but extended 2D solution.
The overall 3D field is constructed by inserting the smaller 3D result into the larger extended 2D result.
This strikes a balance to expend computational resources based on where precision is required.

This optimization of embedding of the smaller 3D solution in the larger extended 2D solution is motivated by observing that the weighting potential reduces rapidly as one considered strips further from the central strip of interest.
For example, Fig~\ref{fig:sl_fr_data} illustrates that the response is down by about 10\% at the second neighbor.

The final major step calculates the induced current of the drifting electron charge.
The drift velocity of the electron at any point is a product of the electron mobility ($\mu$) and the local drift electric field ($\Vec{E}_{drift}$).
The mobility further depends on the liquid argon temperature\cite{bnlurl,microbooneLARP1}:
\begin{equation}
    \Vec{v}_{\rm{drift}}=\mu\cdot \Vec{E}_{\rm{drift}}.
\end{equation}
Given the drift velocity at each grid point, a fourth order Runge-Kutta algorithm, as implemented by the Python SciPy package~\cite{scipy}, is used to calculate fixed steps in time along the drift path of an electron.
Electrons are started on a uniform 2D grid at the domain boundary furthest from the PCB.
This grid is aligned to features of the strips and holes and the grid required by the end application (see section~\ref{sec:valid_procedure}).
Figure \ref{fig:Path} shows example paths after they have propagated toward the strips and through the holes.
Finally, the drift velocity along path steps is included with the weighting field to calculate the induced current on the strip of interest via Eq.~\eqref{eq:SRtheorem}.


Using a single GPU machine (Nvidia 2080Ti), serial calculation of the drift and weighting fields requires five days.
Sufficient GPU resources allow this to be reduced to less than three days by calculating the different fields in parallel. 
The remaining calculation is not GPU accelerated and requires 160 CPU core-minutes.
The software package {\it pochoir} used for these calculations along with installation and usage instructions is available online.~\cite{pochoir}.

\section{Validation of Field Response Calculation}~\label{sec:valid}
\subsection{Wire-Cell Toolkit}
\label{sec:wct}

The calculated FR model was input to the Wire-Cell Toolkit (WCT) \cite{wcweb}.
The two primary components to consume an FR model are the LArTPC detector response simulation and signal processing.
These both provides state-of-art support for wire-based designs but as such they assume translational symmetry along the wire direction and require 2D FR models.

The core operations of these two WCT components are convolutionary and are implemented with the fast Fourier transform (FFT) algorithm.
The transforms are applied in a 2D domain spanning longitudinally along the drift time steps and transversely across the electrodes~\cite{MicroBooNE:2018swd,microbooneLARP2}. 
The detector simulation convolutions are defined at a finer scale than that of the deconvolution in the signal processing and thus two FR models must be prepared to match.
The simulation accounts for sub-strip field variations at points indicated by the red stars in Fig.~\ref{fig:Path}.
On the other hand, the signal processing consumes information provided by the detector readout that represents an averaging over these sub-strip variations.
The only field variation  to remain in the input to the signal processing is at inter-electrode scale.
Thus, the FR model provided to signal processing is formed by averaging sub-strip paths of the simulation FR model. The longer-range variation is retained by this FR spanning a total of 21 strips.

As we see presence of variation along the strip directions (section~\ref{sec:3D_var}) in our simulation of this PCB-based design, there is not yet evidence of these variations in real data.
If improvements allow such variation to be resolved there will be a strong motivation to produce a simulation that utilizes a 3D FR model.
Likewise, given an initial reconstruction of 3D ionization pattern it is possible to utilize a 3D FR model for a more accurate form of signal processing
The WCT developers have investigated providing both of these advancements but have deferred the development as the required computational and software development resources are unavailable.

In the next section, we will use the tools provided in the WCT to validate the field response calculation by comparing simulated and real data from the CERN 50-L prototype detector both before and after signal processing is applied.

\subsection{Procedure}
\label{sec:valid_procedure}

The digital waveform signal from each the electronics channel of the detector is simulated, in part, as a convolution of ionization electron distribution, field response, and electronics response.
The signal processing represents an approximate reversal of this convolution resulting in a sampled estimate of ionization electron distribution as it arrives at the readout electrodes.
Thus the Wire-Cell Toolkit components providing these operations along with data from the real detector provide tools to validate the calculated field response functions. 

The first validation procedure begins by selecting and categorizing readouts from the real detector as described in section~\ref{sec:datasets} and equivalent simulated readouts are prepared.
The waveforms in a given readout are offset in time so that their samples with maximum ADC become mutually aligned.
The multiple aligned readouts from a given category are averaged.
In some cases the aligned waveforms are further summed across the channels.
The averaging and summing remove fluctuations unrelated to the field response modeling accuracy and allow more precise comparisons between simulated and real detector data.



The second validation procedure involves applying the signal processing to the simulated and real detector readouts.
Ideally, the signal processing produces two independent estimates of the same ionization electron charge distributions from the two planes.
We thus compare total charge between the two planes and between simulated and real detectors.


\subsection{Data sets}
\label{sec:datasets}

We apply the validation procedures described above to datasets formed by selecting from the real or simulated readouts that contain activity either from the products of $^{39}$Ar beta decay in the detector or from cosmic-ray muons traversing the detector.
These two selections probe the FR in complimentary ways.

An $^{39}$Ar beta decay produces an electron with energy distributed up to an endpoint of $\approx$0.56~MeV which results in an ionization track no longer than $\approx$0.6~mm, far smaller than strip widths.
Effectively, this results in a point-like source of ionization electrons and their readout directly represents an approximate measure of the field response.
This measure however is smeared by drift effects, averaged over variations at the strip and sample time level and shaped by the detector channel amplifier and ADC.

\begin{figure}[th]
    \centering
    \begin{subfigure}[b]{0.49\textwidth}
        \centering
        \includegraphics[width=1.0\textwidth]{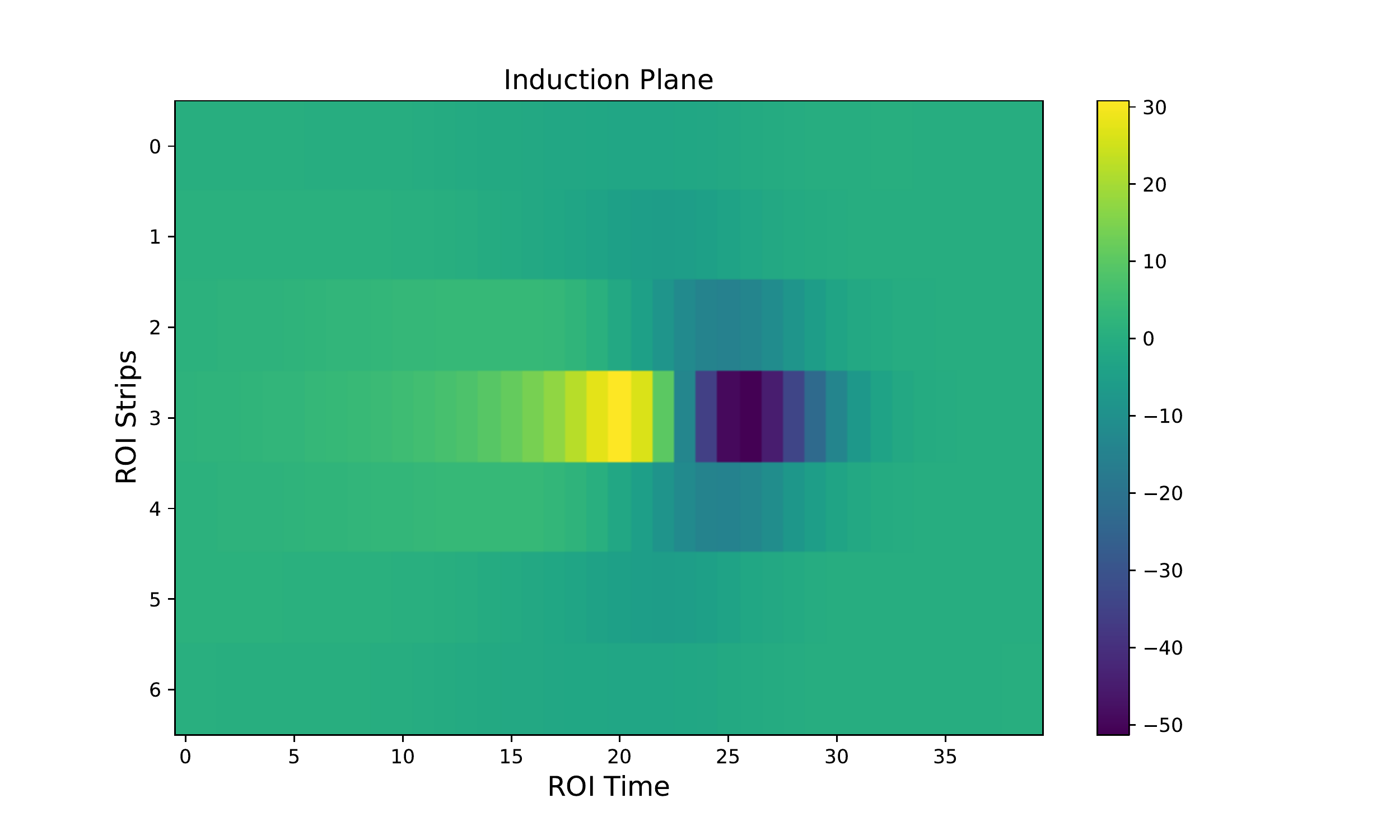}
        \caption{\centering Induction Plane average ROI}
    \label{fig:ROII}
    \end{subfigure}
\hfill    
    \begin{subfigure}[b]{0.49\textwidth}
        \centering
        \includegraphics[width=1.045\textwidth]{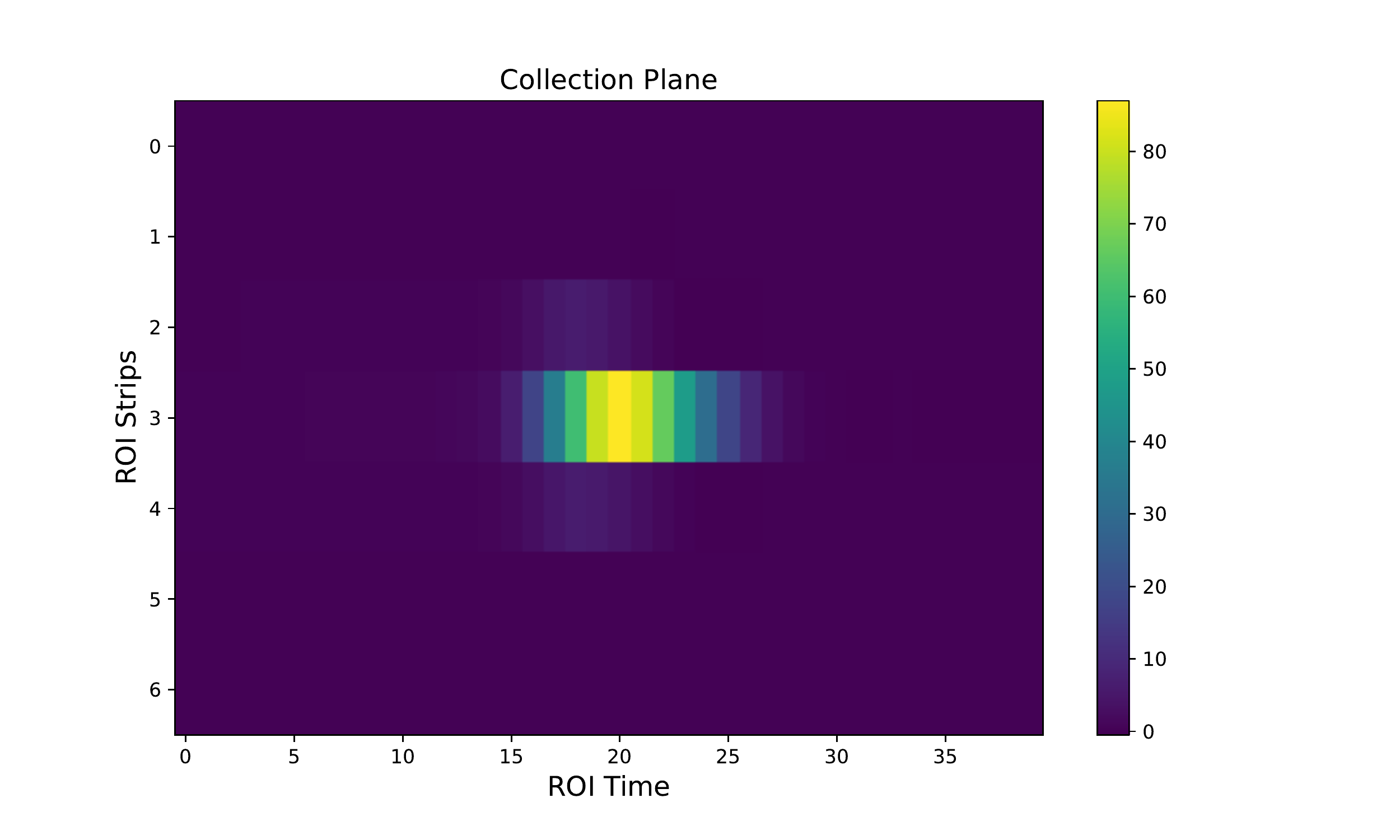}
        \caption{\centering Collection Plane average ROI}
        \label{fig:ROIC}
    \end{subfigure}
  
    \caption{The example of average ROI for $^{39}$Ar data events for 2.5~mm hole PCB region. The signal is averaged across $\sim350$ events. }
    \label{fig:ROI}
\end{figure}

From the real detector readouts rich in  $^{39}$Ar decays are selected by requiring that they contain a single, isolated and localized peak in the 2D ADC channel \textit{vs} sample time array from the collection plane.
This selection was then divided into two subsamples each approximately containing 350 readouts each.
The subsampling was based on whether the peak was in the region of the detector with 2~mm or with 2.5~mm through holes.  A region of interest was then defined to be centered on each peak, extended over $\pm$25 ADC sample periods and covering $\pm$3 channels and the central channel containing the peak.
After the averaging technique described above, the ADC in the region-of-interest for 2.5~mm holes is shown in Fig.~\ref{fig:ROI}. 

On the other hand, the cosmic-ray muons produce long tracks which leads to a variable amount of mixing of longitudinal and transverse field response features depending on track angle.
The muons are approximately minimum ionizing and those that undergo discrete energy loss such as from delta-ray production or bremsstrahlung can be removed from the samples.
Thus, they may provide a source of ionization electrons which is approximately linear and uniform. Consequently, we utilize tracks for studies of field response shapes instead of $^{39}$Ar.

\begin{figure}[th]
  \centering
  \includegraphics[width=1\linewidth]{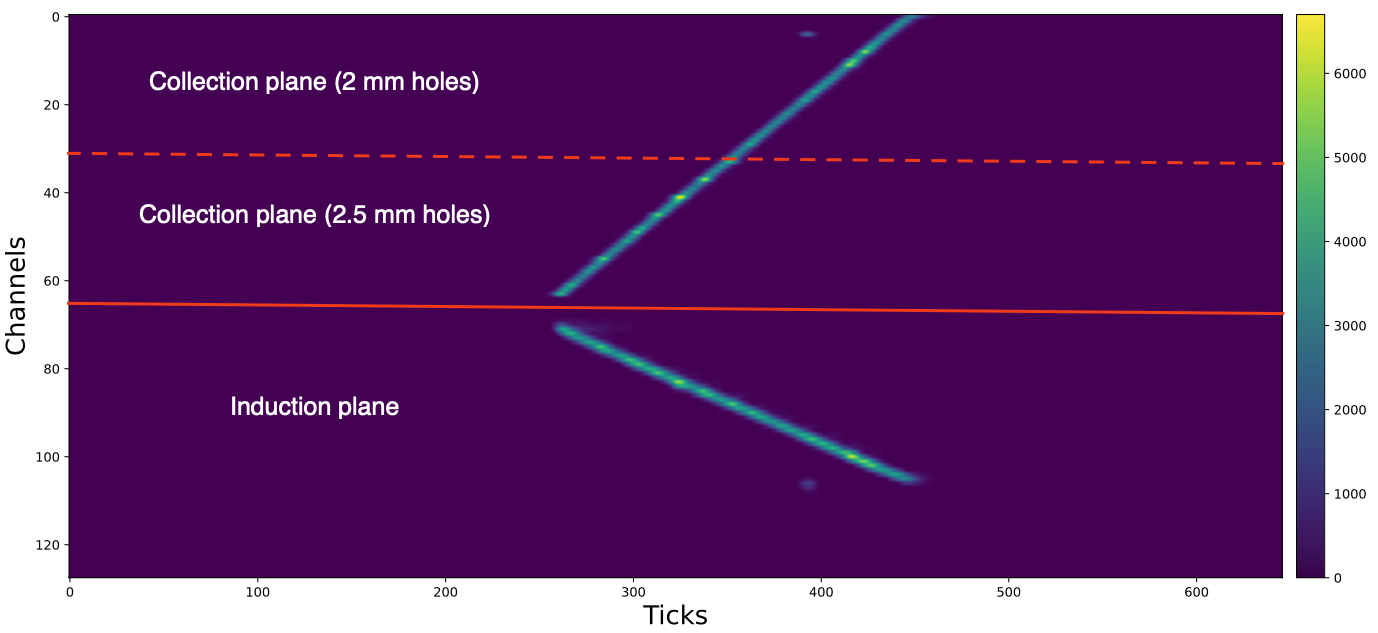}  

    \caption{One example of a cosmic-ray muon track from the CERN 50-L prototype that 
    has $\sim 65^\circ$ azimuth angle and crosses all collection strips.
    The reconstructed ionization electron distributions after the signal processing is shown 
    as a function of the time in tick (i.e. 0.5 $\mu$s). }
    \label{fig:track}
\end{figure}

The selection for cosmic-ray muons required that waveforms from at least 20 consecutive channels of the collection plane provide at least one ADC sample of at least 40 counts.
We then selected most track topologies that were detectable by CERN 50-L prototype design with a final manual examination of events to get a single straight track. This gave tracks directions of $\sim65^\circ$ azimuth angle.
One example is shown in Fig.~\ref{fig:track}.

\subsection{PCB hole sizes}
\label{sec:exp2D}

\begin{figure}[ht]
    \centering
    
    \begin{subfigure}[b]{0.49\textwidth}
        \centering
        \includegraphics[width=1.0\textwidth]{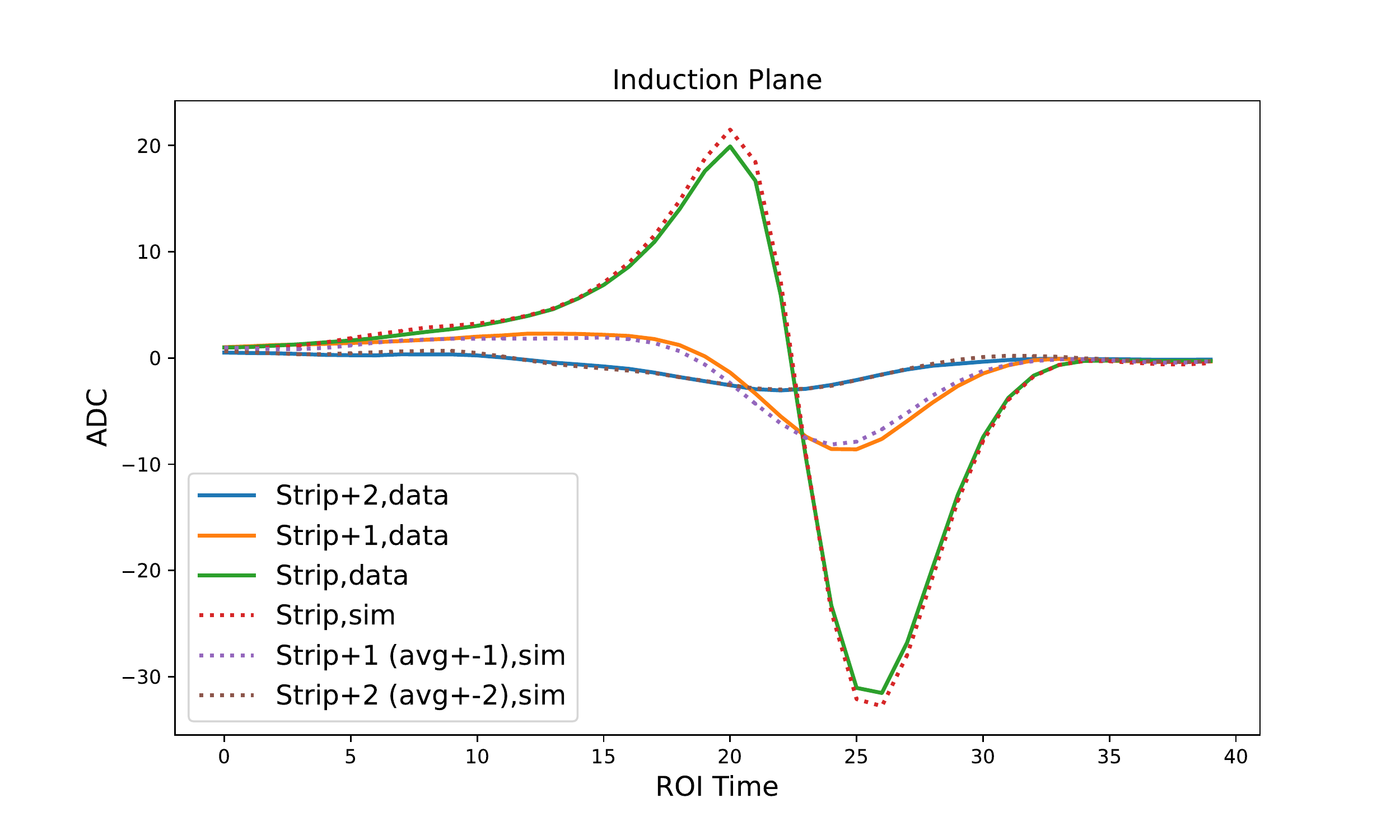}
        \caption{\centering Induction plane average waveform}
        \label{fig:sl_fr_data_I}
    \end{subfigure}
\hfill
    \begin{subfigure}[b]{0.49\textwidth}
        \centering
        \includegraphics[width=1.0\textwidth]{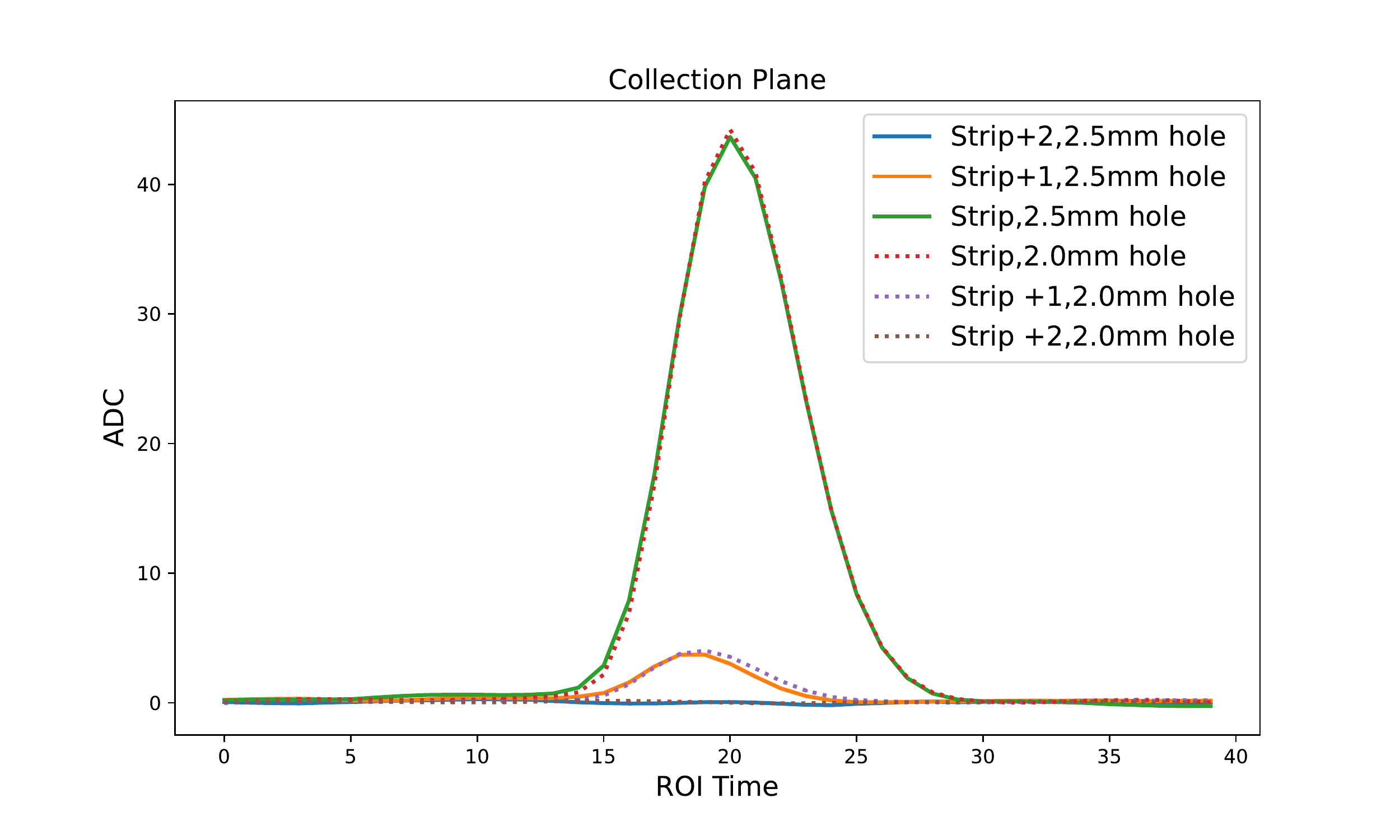}
        \caption{\centering Collection plane average waveform}
        \label{fig:sl_fr_data_C}
    \end{subfigure}

    \caption{The comparison between average waveform extracted from $^{39}$Ar data events for 2.5 mm diameter holes (solid line) 
    and 2.0 mm diameter holes (dashed line). The field response is from the middle strip (with maximum ADC value) and two 
    strips on each side.}
    \label{fig:sl_fr_data}
\end{figure}

With the point-like energy depositions from $^{39}$Ar sample, the effect of hole size on the field response was estimated.
The average waveforms for the central and two nearest strips are shown in Figure \ref{fig:sl_fr_data} for the two subsamples covering regions of different through hole size.
For the collection plane, they indicate that small holes produce $\approx 1\%$ higher peak ADC than do large holes.
An increase of $\approx 3\%$ is observed in the positive peak for the induction plane. 
The largest difference occurs at the middle strip.


\begin{figure}[ht]
    \centering
    
    \begin{subfigure}[b]{0.49\textwidth}
        \centering
        \includegraphics[width=1.0\textwidth]{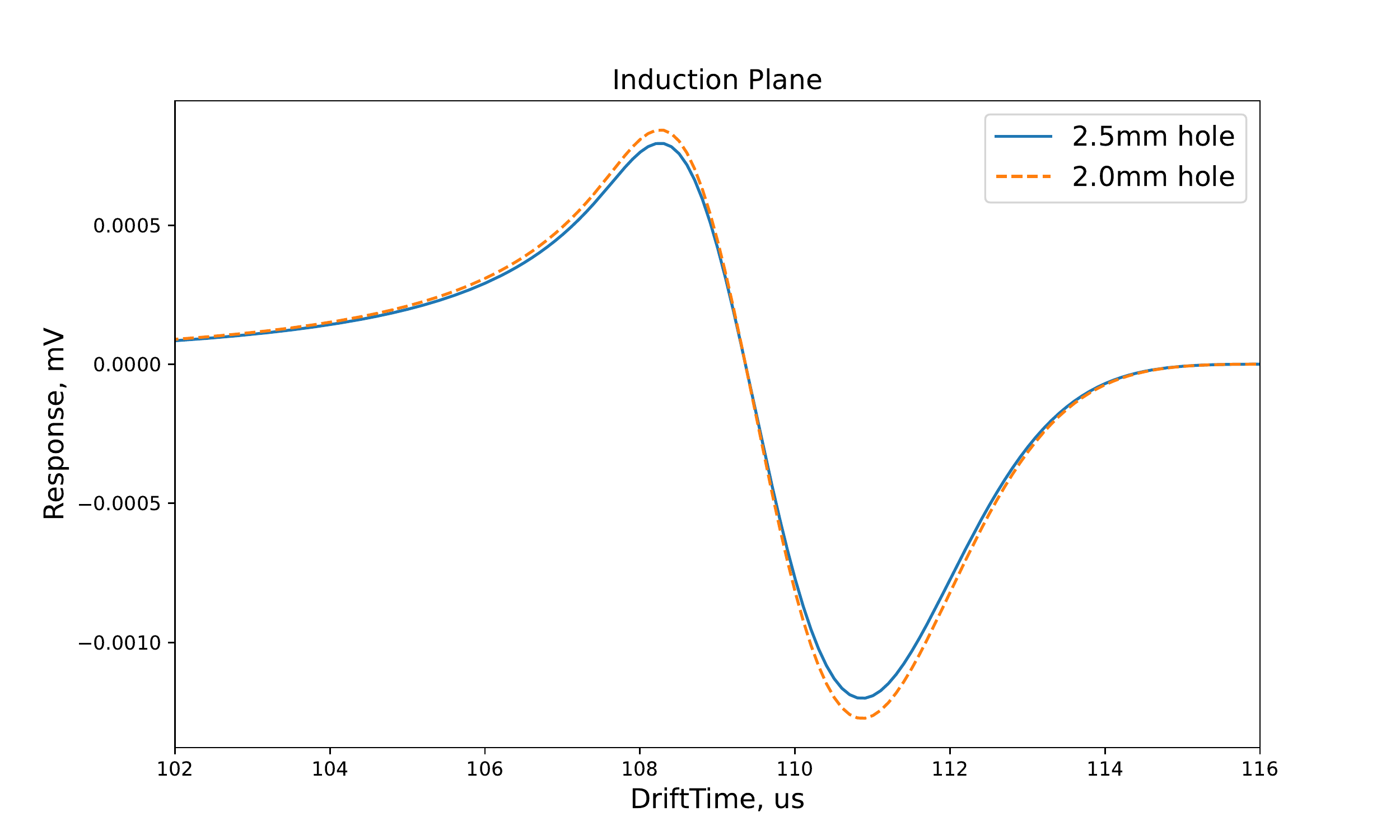}
        \caption{\centering Induction plane average waveform}
        \label{fig:sl_fr_mc_I}
    \end{subfigure}
\hfill
    \begin{subfigure}[b]{0.49\textwidth}
        \centering
        \includegraphics[width=1.0\textwidth]{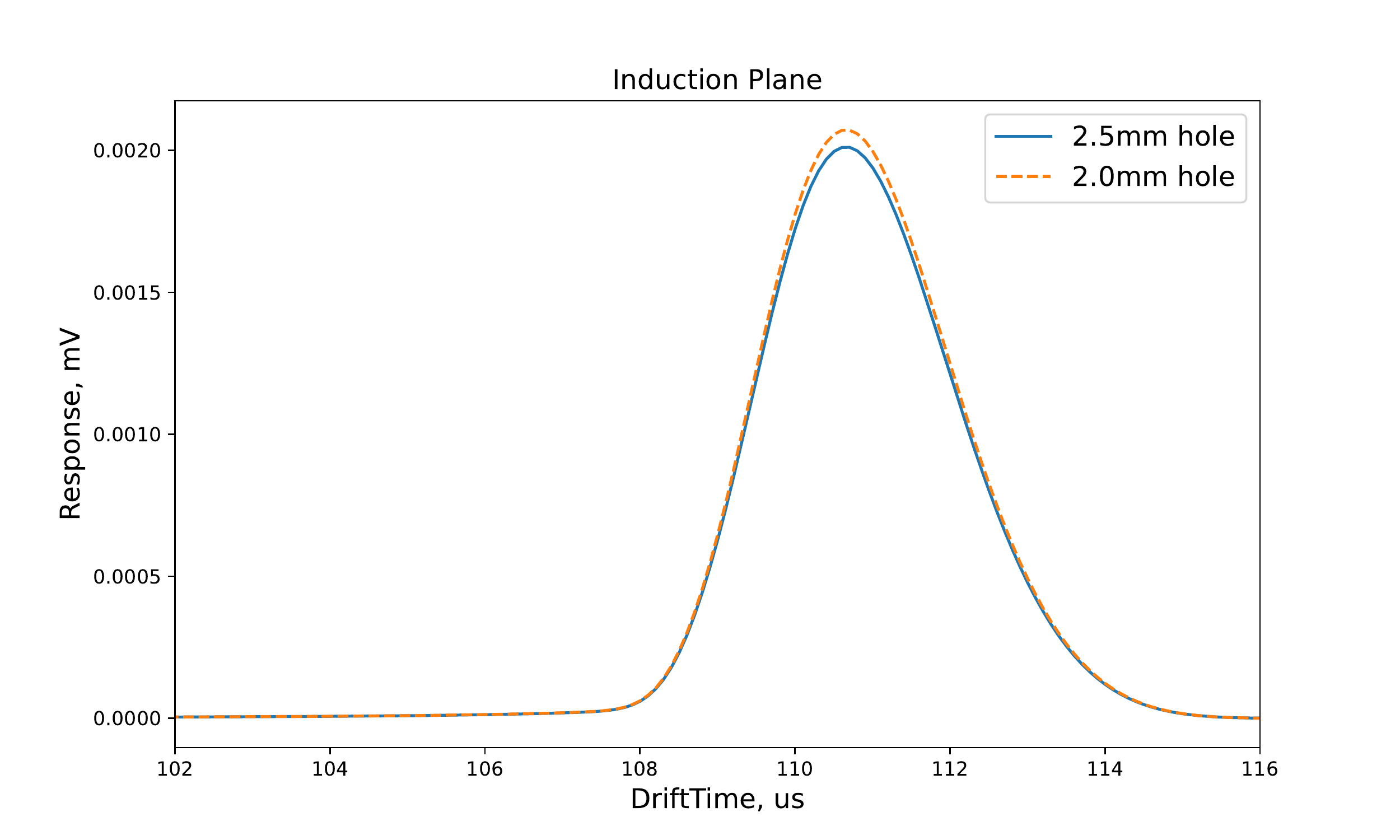}
        \caption{\centering Collection plane average waveform}
        \label{fig:sl_fr_mc_C}
    \end{subfigure}

    \caption{The comparison between simulated average waveform for 2.5 mm diameter holes (solid line) and 
    2.0 mm diameter holes (dashed line) in the middle strip (with maximum voltage). }
    \label{fig:sl_fr_mc}
\end{figure}

We also check if the effect for the middle strip can be reproduced in the simulation.
We carried out the field response
calculations for both 2.5 mm and 2.0 mm hole sizes.
The $^{39}$Ar events are simulated 
as point sources for simplicity and placed  centered on a single strip.  
Figure \ref{fig:sl_fr_mc} shows the convolution of field and electronics response.
It shows the increase in signal for small holes compared to large ones seen in real detector data is present in the simulation.
The two signal peaks are different in the simulation by about $3\%$ and $5\%$ for collection and induction planes, respectively. 

\subsection{Cosmic-ray muon tracks}

As discussed in Sec.~\ref{sec:valid_procedure}, use of cosmic-ray muons enable complimentary tests of the field response and in particular ones that are sensitive to admixtures of longitudinal and transverse long-range induction effects.
The long-range induction components are generally more predominantly asserted in induction plane signals as the collection plane is shielded from long range fields by the induction plane.
For this validation, we select specific cosmic-ray muons at $\sim65^\circ$ azimuth angle.
Their equivalent data is simulated using straight line tracks with an energy deposition of 2.5 MeV/cm as a simplifying proxy for minimum ionizing muons. 

\begin{figure}[ht]
    \centering
    
    \begin{subfigure}[b]{0.49\textwidth}
        \centering
        \includegraphics[width=1.0\textwidth]{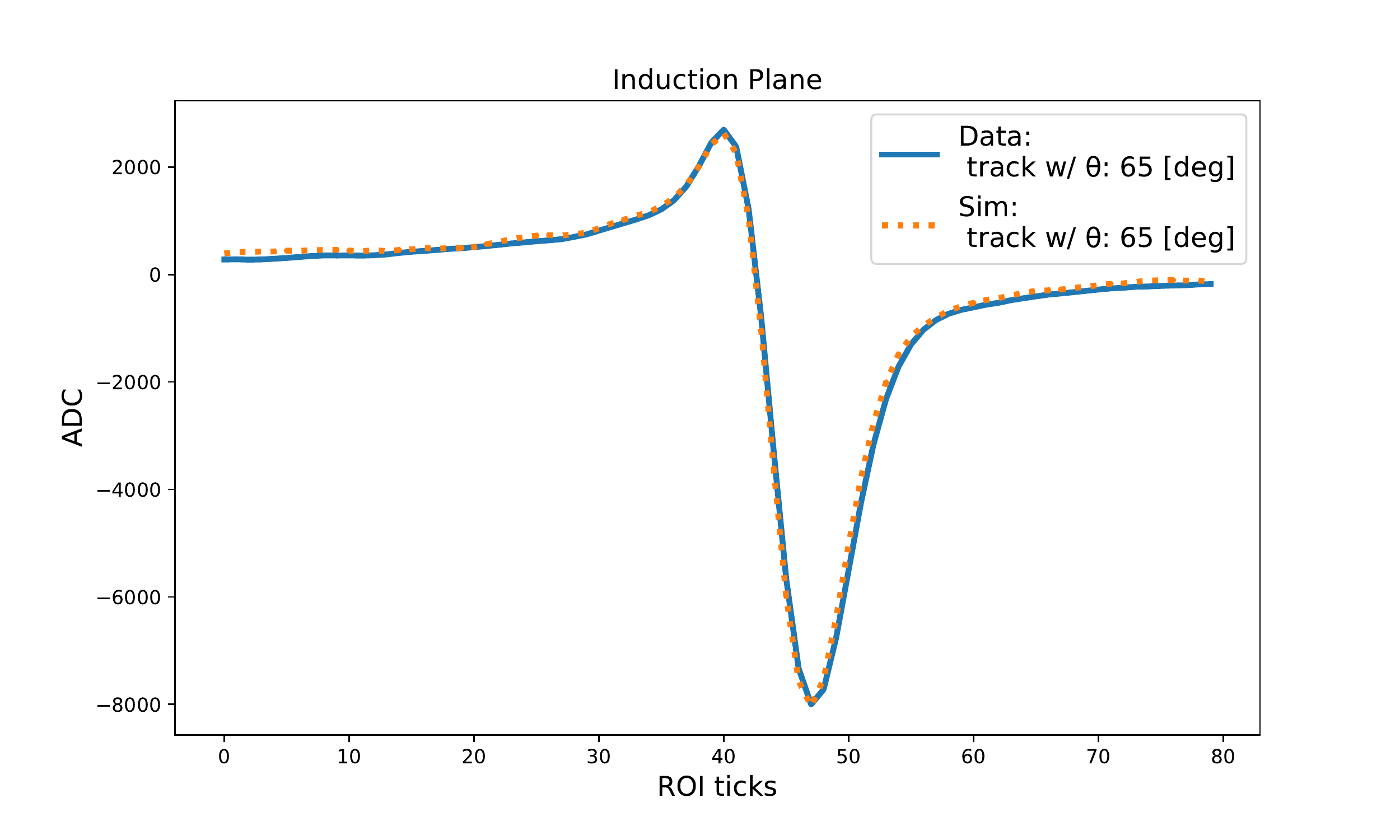}
        \caption{\centering Induction plane field response}
        \label{fig:track_fr_I}
    \end{subfigure}
\hfill
    \begin{subfigure}[b]{0.49\textwidth}
        \centering
        \includegraphics[width=1.0\textwidth]{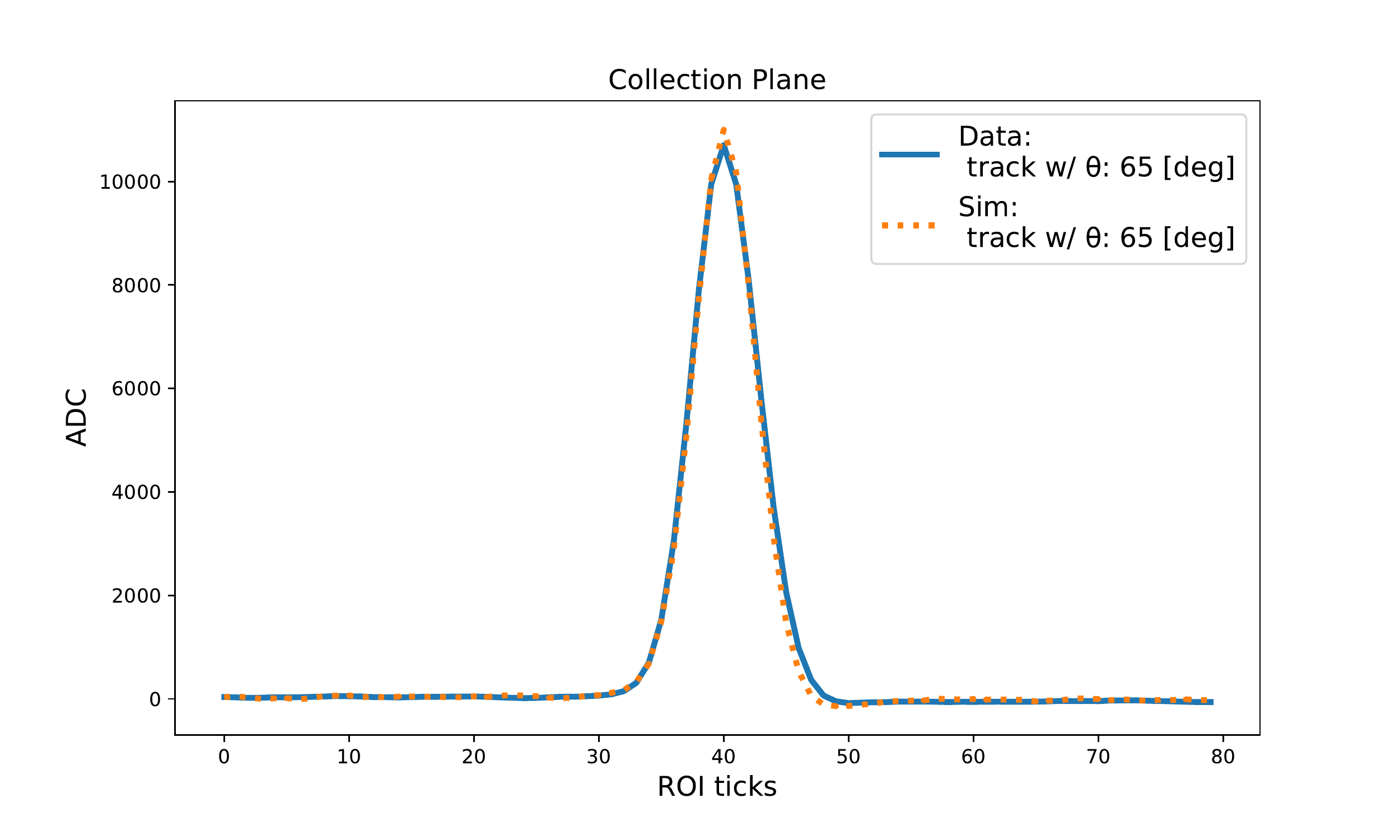}
        \caption{\centering Collection plane field response}
        \label{fig:track_fr_C}
    \end{subfigure}

    \caption{The comparison of the summed waveform after peak alignment between data and simulation for 
    cosmic-ray muon track with topology shown in Fig.~\ref{fig:track}. The data and simulation are area-normalized. 
    The normalization factors are consistent within 4\% between collection and induction planes.}
    \label{fig:track_fr}
\end{figure}

Figure \ref{fig:track_fr} shows the comparison of summed waveforms between simulated
cosmic-ray muon data and real detector for both induction and collection planes.
In order to examine the shape, the area of simulation is adjusted to match that of the data. 
The normalization factors are consistent within 4\% between collection and induction planes.
The remaining difference may come from the presence of both hole sizes in the data as well as hardware effects (e.g. bias voltage capacitor as discussed in Sec.~\ref{sec:detector}) that are omitted in the simulation.

\begin{figure}[ht]
    \centering
    
    \begin{subfigure}[b]{0.49\textwidth}
        \centering
        \includegraphics[width=1.0\textwidth]{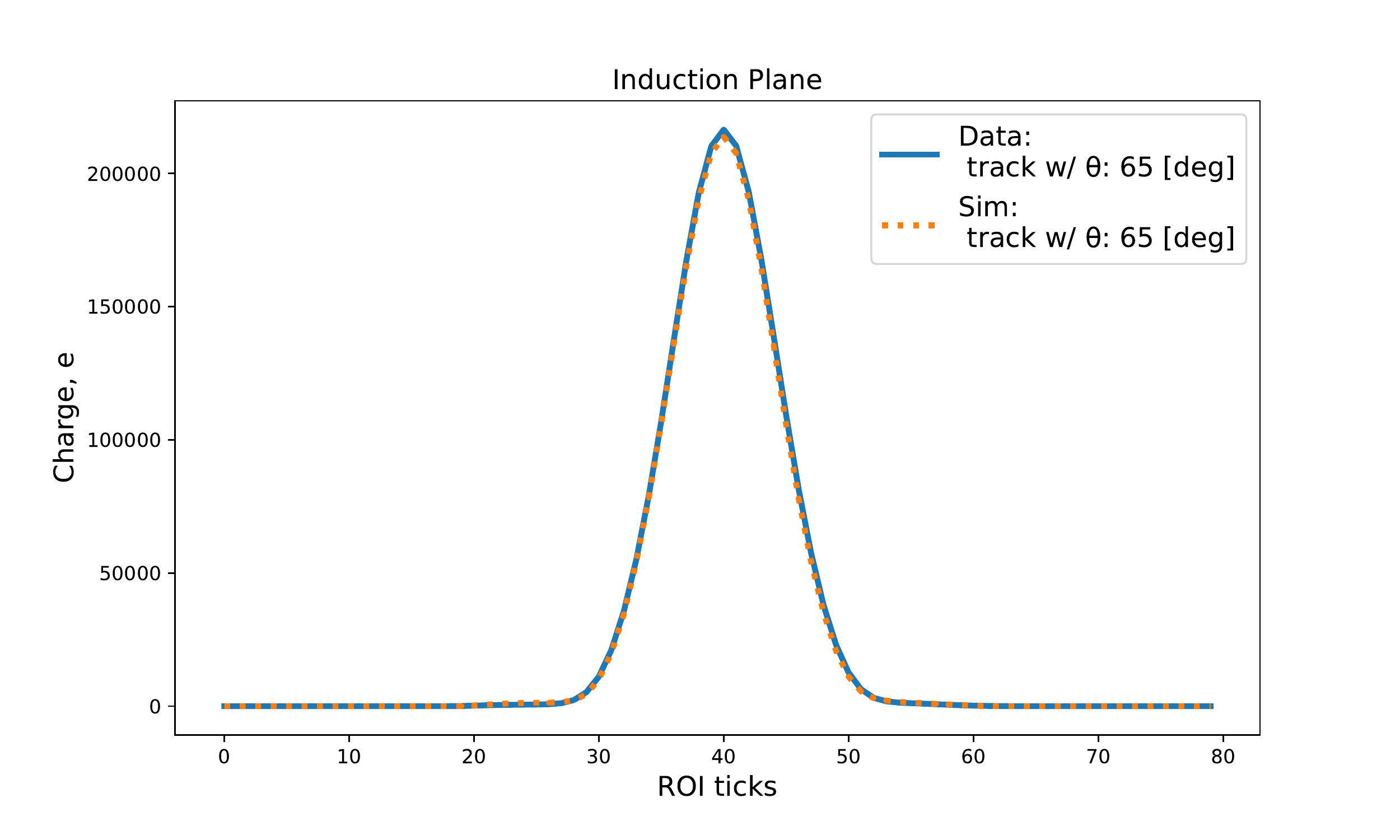}
        \caption{\centering Induction plane charge}
        \label{fig:track_ch_I}
    \end{subfigure}
\hfill
    \begin{subfigure}[b]{0.49\textwidth}
        \centering
        \includegraphics[width=1.0\textwidth]{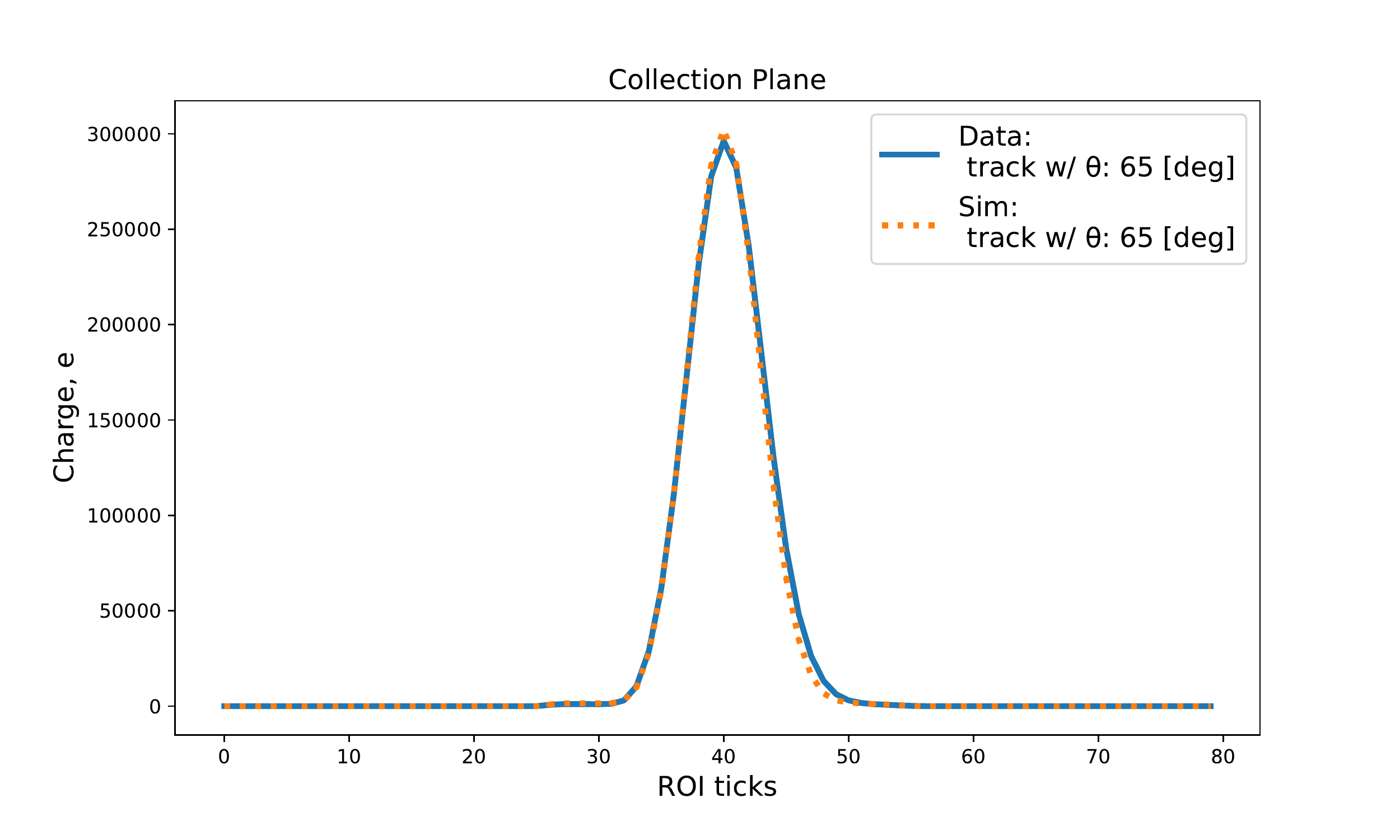}
        \caption{\centering Collection plane charge}
        \label{fig:track_ch_C}
    \end{subfigure}
    \caption{The comparison of the reconstructed ionization charge distribution between data and simulation for 
    cosmic-ray muon track with topology shown in Fig.~\ref{fig:track}. The data and simulation are area-normalized.
    The normalization factors are consistent within 4\% between collection and induction planes.}
    \label{fig:track_ch}
\end{figure}

Figure \ref{fig:track_ch} shows the comparison of reconstructed ionization charge distribution after the signal 
processing between data and simulation. Similar to the comparison in raw waveform, the simulation results are also 
area normalized to those of data separately for induction and collection planes. The shapes after the normalization 
show a good agreement between data and simulation. The normalization factors are consistent within 4\% between 
collection and induction planes. Therefore, the simulated overall detector response is shown to be consistent to 
data from a Vertical Drift (VD) detector prototype at CERN within 5\% in central values

\begin{figure}[ht]
    \centering
    
    \begin{subfigure}[b]{0.46\textwidth}
        \centering
        \includegraphics[width=1.0\textwidth]{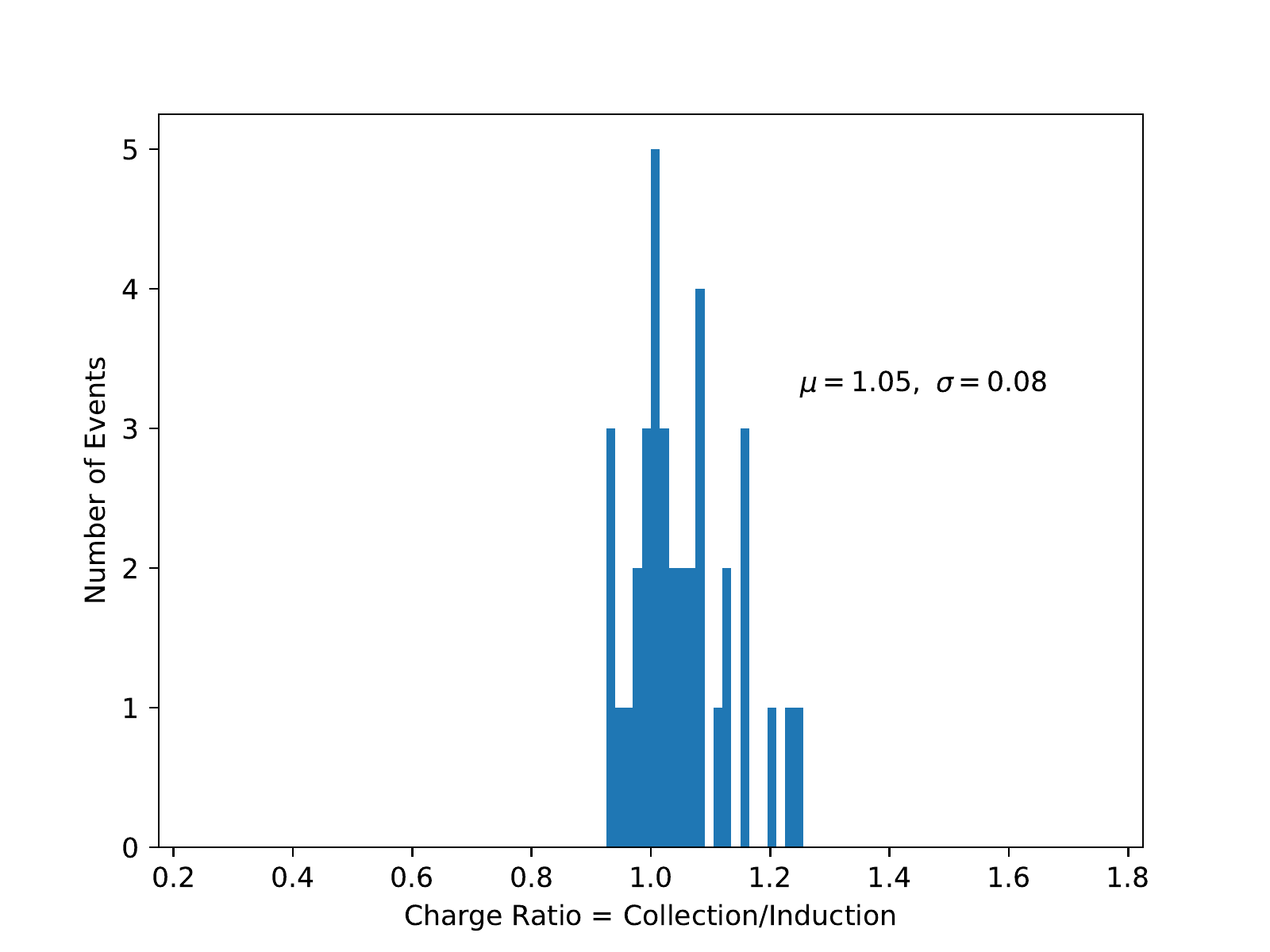}
        \caption{\centering Data cosmic tracks}
        \label{fig:ratio_track_data}
    \end{subfigure}
\hfill
    \begin{subfigure}[b]{0.5\textwidth}
        \centering
        \includegraphics[width=1.0\textwidth]{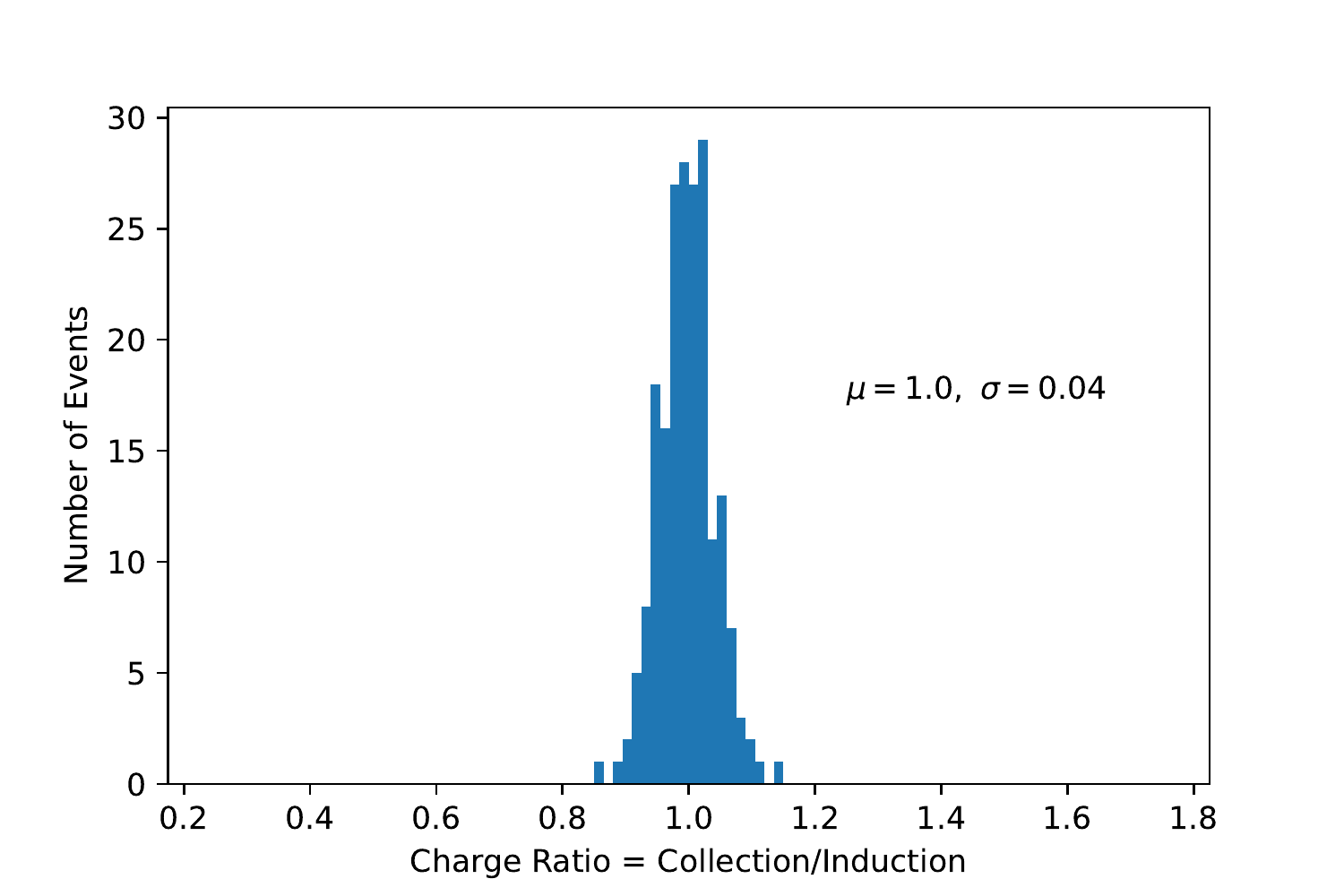}
        \caption{\centering Simulated cosmic tracks}
        \label{fig:ratio_track_sim}
    \end{subfigure}

    \caption{The ratio of total charge seen by collection and induction planes from data and simulated cosmic-ray 
    muon tracks. The RMS sigma is used as a result of limited statistics for data tracks.}
    \label{fig:ratio_track}
\end{figure}

As introduced in Sec.~\ref{sec:valid_procedure}, the comparison of the reconstructed charge between 
collection and induction plane can also be used to validate the overall field response functions. 
Figure \ref{fig:ratio_track_data} shows the ratio of the charge measured by the collection plane
to that measured by the induction plane for a set of cosmic-ray muon tracks, which includes strict selection of straight tracks that cross multiple collection and 
induction strips. We further discard tracks that cross the anode plane, since they would introduce distortion in the 
reconstructed charge at the beginning of the track. 
In comparison, Fig.~\ref{fig:ratio_track_sim}, shows the ratio between reconstructed charge from collection 
and induction planes for simulated cosmic-ray muon tracks (i.e. straight lines) with similar track angles. The 
variation of this ratio in the simulation is smaller than that of the data. This difference may partially come 
from the fact that the selection does not differentiate between different hole sizes.
While some tracks can fully or partially cross only 2 mm holes, the field response functions used in the 
simulation are calculated assuming 2.5 mm diameter holes.
In addition, the variation of the field response functions
along the strip, which was not considered in the 2D detector response calculation, may play a role. 

\subsection{Variations along the strip}\label{sec:3D_var}

\begin{figure}[ht]
    \centering
    
    \begin{subfigure}[b]{0.49\textwidth}
        \centering
        \includegraphics[width=1.0\textwidth]{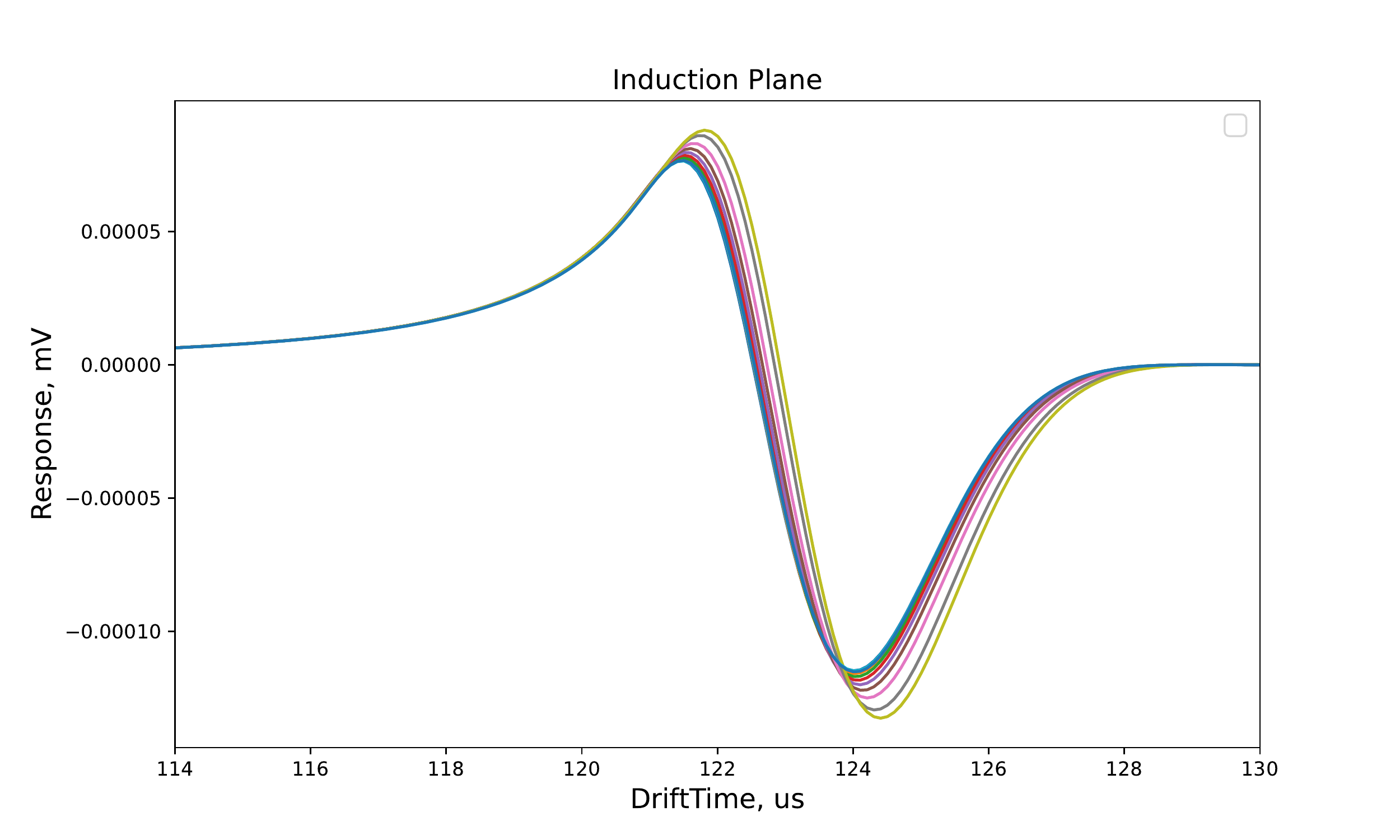}
        \caption{\centering Induction plane waveform}
        \label{fig:stripdir_fr_I}
    \end{subfigure}
\hfill
    \begin{subfigure}[b]{0.49\textwidth}
        \centering
        \includegraphics[width=1.0\textwidth]{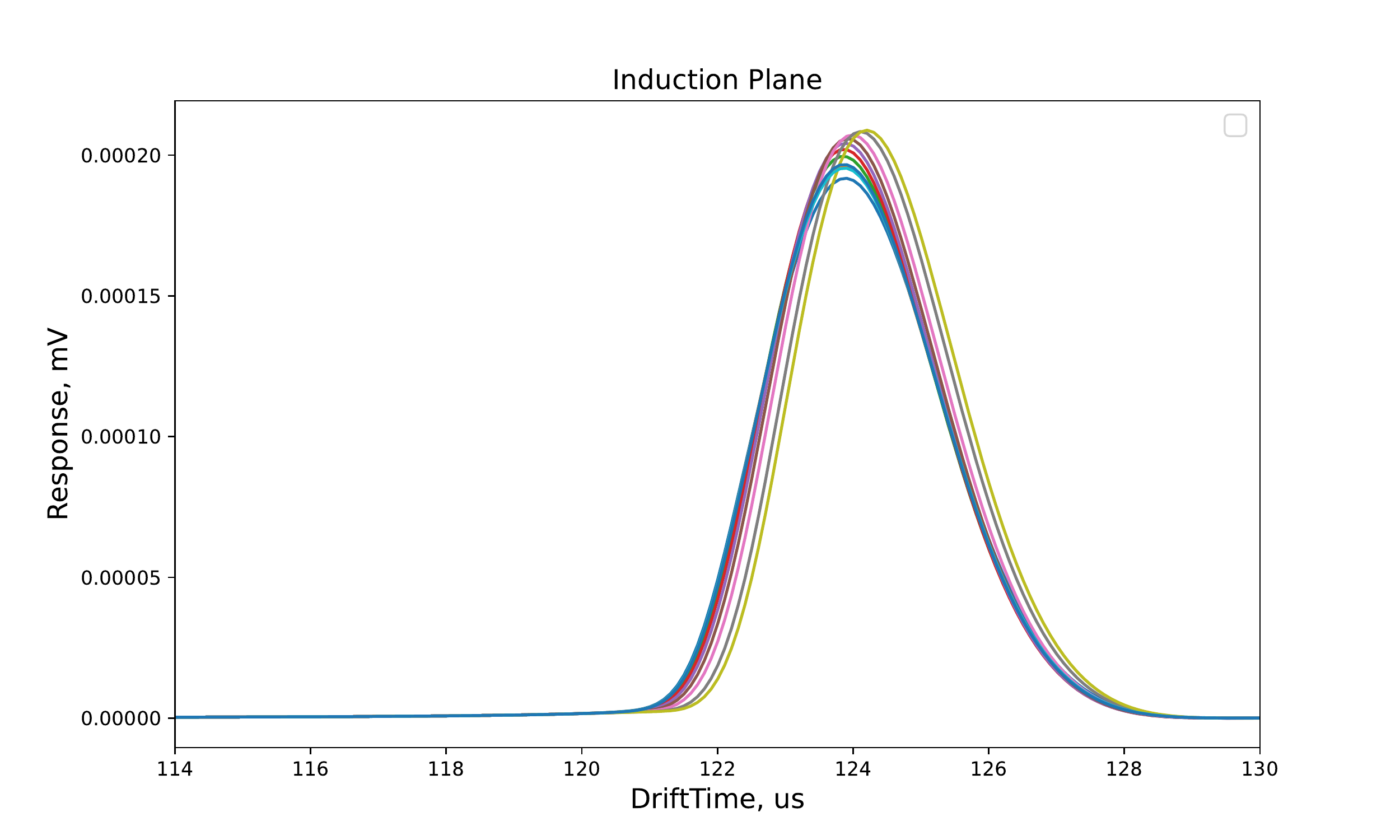}
        \caption{\centering Collection plane waveform}
        \label{fig:stripdir_fr_C}
    \end{subfigure}

    \caption{The simulated waveform using different field response functions for induction and collection planes. 
    Different colors represent the current for different paths along the strip orientation shown in Fig.~\ref{fig:Path}.
    }
    \label{fig:stripdir_fr}
\end{figure}

As described in Sec.~\ref{sec:fralg}, the FR is modeled in mix of 2D and 3D domains and then averaged to a 2D model for use by WCT~\cite{wcweb} detector response simulation.
In order to estimate 
the impact of this approximation, we perform $^{39}$Ar simulations with field response functions at each of the 
11 paths along the strip (shown in Figure \ref{fig:Path}).
The convolutions of a point source with the field response along each path and with the electronics response are shown in Fig. \ref{fig:stripdir_fr}.
They illustrate how a charge drifting at different positions along the strip may indeed induce differently shaped response in the detector and thus violate translational symmetry.
While the collection plane signal varies by $4\%$ in terms of the peak height, the induction signal variation is 
at $\sim12\%$ from the average field response. 

This effect is translated to reconstructed charge by simulating a point source in the detector with the nominal 2D field response model but reconstructing its charge with a variety of calculated responses at different specific positions along the strip orientation. 
This uncovers a $\sim 7\%$ variation of reconstructed charge on induction plane and negligible variation on collection plane. 
When considering the track topology and charge diffusion, the impact from this fine-scale position dependence along the strip orientation is expected to be smaller than 7\%.

\begin{figure}[ht]
    \centering
    
    \begin{subfigure}[b]{0.49\textwidth}
        \centering
        \includegraphics[width=1.0\textwidth]{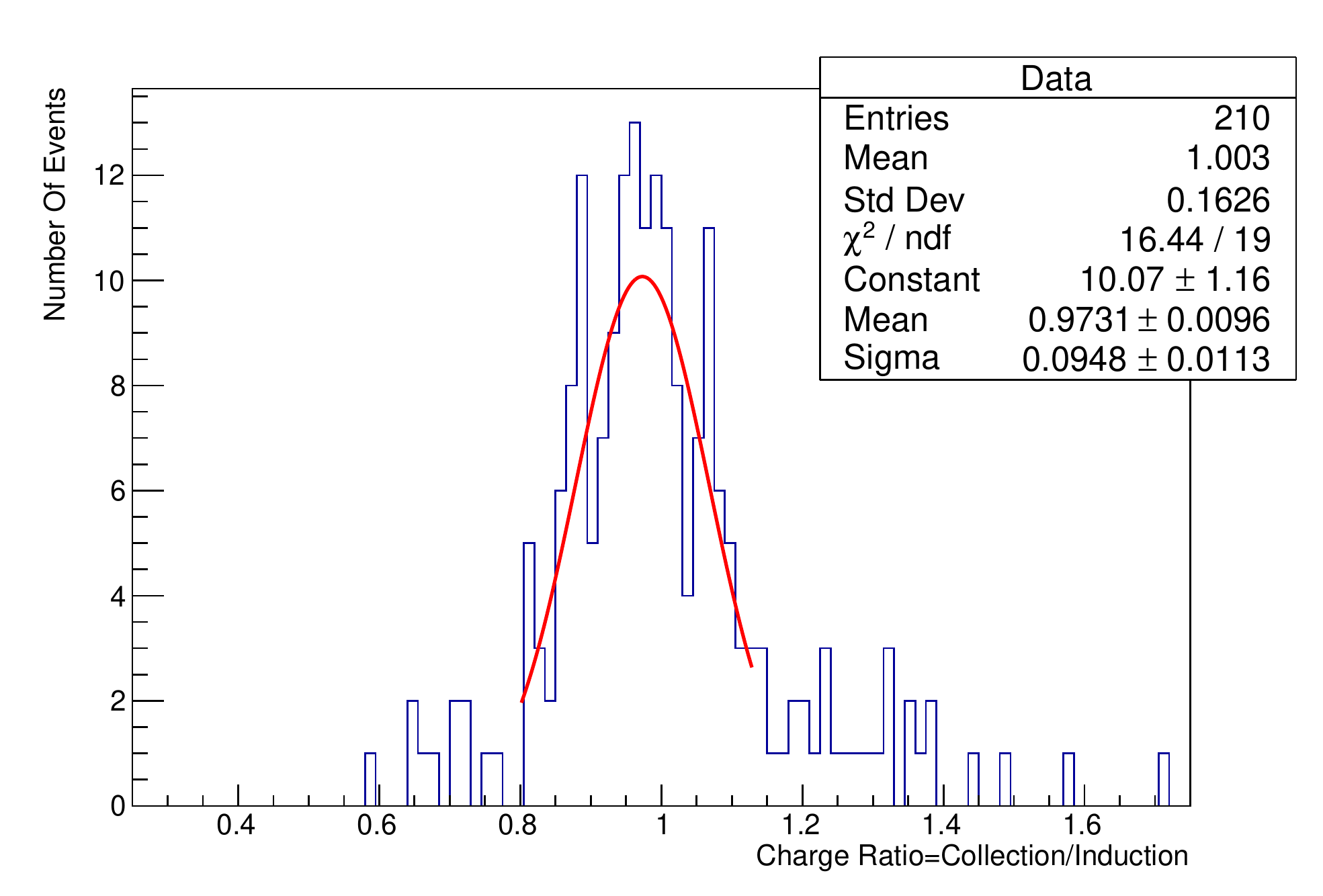}
        \caption{\centering Data $^{39}$Ar events}
        \label{fig:ratio_ar_data}
    \end{subfigure}
\hfill
    \begin{subfigure}[b]{0.49\textwidth}
        \centering
        \includegraphics[width=1.0\textwidth]{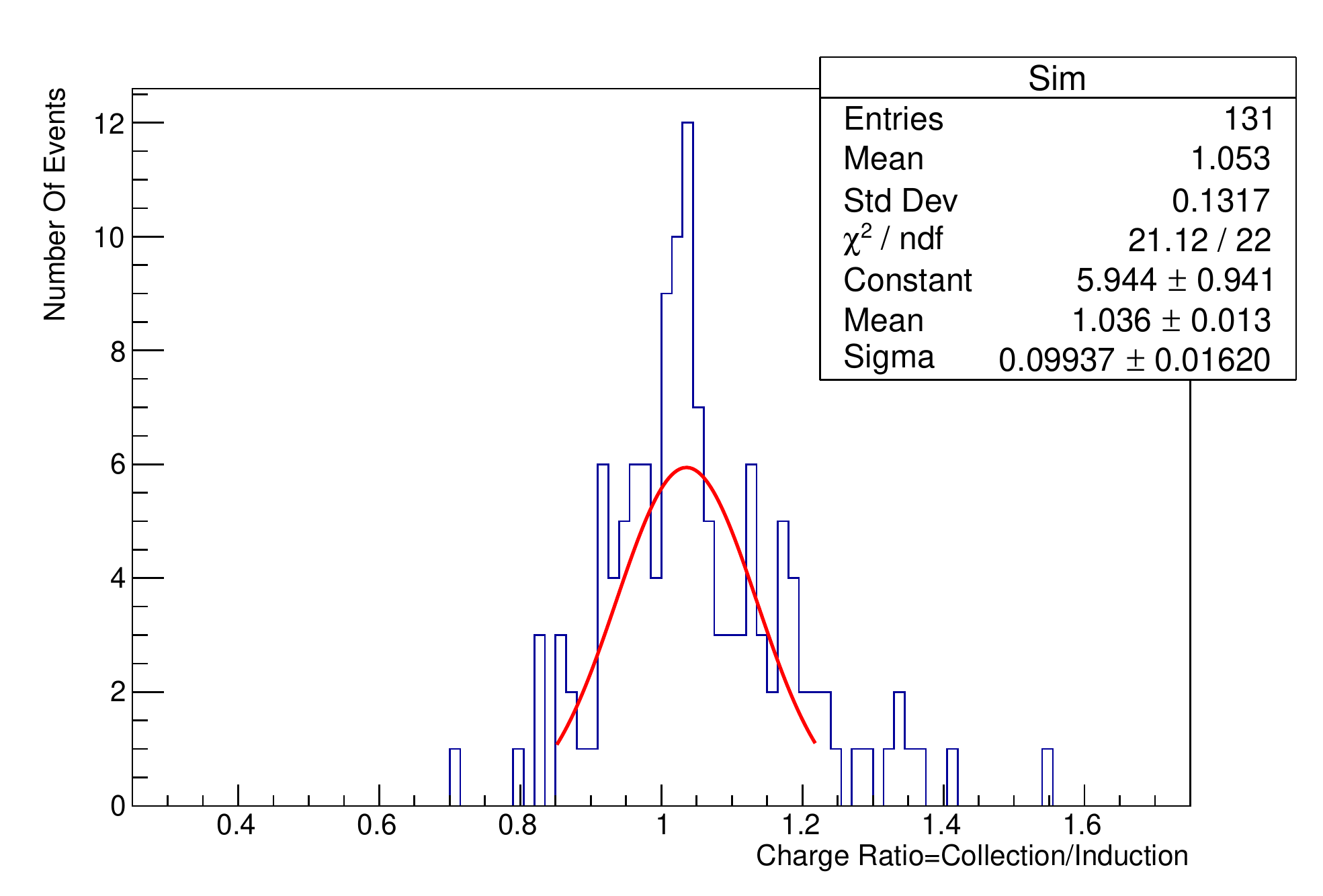}
        \caption{\centering Simulated $^{39}$Ar events}
        \label{fig:ratio_ar_sim}
    \end{subfigure}
    \caption{The ratio of total charge seen by collection and induction planes from data and simulated $^{39}$Ar events.}
    \label{fig:ratio_ar}
\end{figure}

In order to further examine this effect, we chose $^{39}$Ar events that occurred on the side of the detector with 2.5 mm holes. 
Since the average 2D field response along the strip is used in the simulation, the ratio of the reconstructed charge 
on the collection-plane channels to that on the induction-plane channels is expected to be wider in data, where the 
ratio would vary depending on the position of $^{39}$Ar along the strip. Figure~\ref{fig:ratio_ar} shows these 
ratios in data and simulation. While the distribution in data has a larger standard deviation (RMS) than that of 
simulation, the best-fit sigmas after fitting the peak region of the distribution with a Gaussian function are 
similar. This observation may be explained by the large intrinsic width in this ratio from the $^{39}$Ar samples 
(e.g. electronics noise, position-dependent field response perpendicular to the strip orientation), so that 
the $\sim$7\% position-dependent variations along the strip orientation is too small to be observed.


    


\section{Summary}
\label{sec:Summ}

In this paper, we present a new hybrid 3D/2D field response computation package {\it pochoir}~\cite{pochoir} that calculates field response for newly proposed printed circuit board (PCB) anodes for LArTPCs.
Utilizing samples of cosmic-ray muons and $^{39}$Ar decays and a detector simulation package assuming 2D symmetry, the simulated field response functions are validated using data from the CERN 50-L prototype detector. The simulated overall detector response is shown to be consistent with data from this Vertical Drift detector prototype within 5\% in the central value.
Position-dependent variations along the strip orientation arising from the hole pattern are not included in the the simulation and this may result in inaccuracy smaller than 7\%.
While the current comparison between simulated and real detector data does not reveal this level of variation, this effect should not be ignored as it may be resolved with more advanced reconstruction, particularly those based on the training of deep neural networks.
Further development of a 3D detector simulation and/or approaches to include this effect in estimating the overall detector systematic uncertainties is a focus of future work with the understanding that they will likely incur large computational and software development costs.  

\bibliography{mybib}

\end{document}